\documentclass[12pt]{article}
\usepackage{amsfonts, amssymb, amscd}
\usepackage{amsmath}
\usepackage[subnum]{cases}
\usepackage{color}
\usepackage{cite}

\def\l{\left}
\def\r{\right}
\def\bl{\Biggl}
\def\br{\Biggr}
\def\nn{\nonumber}
\def\ql{\textquotedblleft}

\def\1o2{{1\over2}}

\def\a{\alpha}
\def\b{\beta}
\def\ab{{\alpha\beta}}
\def\g{\gamma}

\def\d{\delta}

\def\m{\mu}
\def\n{\nu}
\def\mn{{\mu\nu}}
\def\k{\kappa}
\def\la{\lambda}
\def\La{\Lambda}
\def\p{\phi}
\def\pa{\partial}
\def\s{\sigma}

\begin{document}

\title{Kerr-Schild-Kundt Metrics in Generic Einstein-Maxwell Theories}
\author{ Metin G\"{u}rses$^{(a)}$\footnote{gurses@fen.bilkent.edu.tr}, Yaghoub Heydarzade$^{(a)}$\footnote{yheydarzade@bilkent.edu.tr}\,\, and \c{C}etin
\c{S}ent\"{u}rk$^{(b)}$\footnote{csenturk@thk.edu.tr} \\
{\small (a) Department of Mathematics, Faculty of Sciences}\\
{\small Bilkent University, 06800 Ankara, Turkey}\\
{\small (b) Department of Aeronautical Engineering}\\
{\small University of Turkish Aeronautical Association, 06790 Ankara, Turkey}
}


\begin{titlepage}
\maketitle
\thispagestyle{empty}

\begin{abstract}
We study Kerr-Schild-Kundt class of metrics in generic gravity theories with Maxwell's field. We prove that these metrics linearize and simplify the field equations of generic gravity theories with Maxwell's field.

\end{abstract}

\end{titlepage}


\setcounter{page}{2}

\section{Introduction}

In the last decade, in a series of papers \cite{ggst,gst1,ghst,gst2,gst3,gst4}, we showed that the Kerr-Schild-Kundt(KSK) types of metrics are universal. This means that the KSK metrics reduce the field equations of any generic gravity theory to a linear equation for the metric function $V(x)$ (see below). By using this result, we have studied some special cases, such as quadratic gravity, f(Riemann)-gravity, cubic gravity theories, and found AdS-plane and pp-wave solutions of these theories. For the universality and almost universality of the KSK metrics see also the works in \cite{hervik1,hervik2,pravda1,pravda2,pravda3,Kuchn}

The KSK metrics are defined by the spacetime metric
\begin{equation}\label{KS}
  g_{\m\n}=\bar{g}_{\m\n}+2Vl_\m l_\n,
\end{equation}
which is in the \ql generalized" Kerr-Schild form \cite{ks,gg}. Here, $\bar{g}_{\m\n}$ represents the background spacetime, $V(x)$ is a scalar field called the profile function, and $l^\m$ is a null vector field. The background metric $\bar{g}_{\m\n}$ is assumed to be maximally symmetric and as such its Riemann tensor satisfies the following property:
\begin{equation}\label{Rback}
  \bar{R}_{\m\a\n\b}=K (\bar{g}_{\mu \alpha} \bar{g}_{\nu \beta}-\bar{g}_{\nu \alpha} \bar{g}_{\mu \beta}),~~~~  K=\frac{\bar{R}}{D(D-1)}=const.,
\end{equation}
where $K$ is the curvature constant and it is related to the background Ricci scalar $\bar{R}$ and the spacetime dimension $D$, as seen. Therefore, depending on the value of $K$, the background might be either the Minkowski, de Sitter (dS), or anti-de Sitter (AdS) spacetime for which $K=0$, $K>0$, or $K<0$, respectively. The profile function $V(x)$ and the vector field $l^\m$ in (\ref{KS}) together satisfy the relations
\begin{eqnarray}
&&l_\m l^\m=0,~~\nabla_\m l_\n=\frac{1}{2}(l_\m \xi_\n+l_\n \xi_\m),\label{lxi}\\
&&l_\m \xi^\m=0,~~l^\m\pa_\m V=0,\label{lV}
\end{eqnarray}
with $\xi^\m$ being an arbitrary vector field which becomes explicit for a specific background metric. With these properties, one can show that the inverse metric $g^\mn$, the Einstein tensor $G_\mn$, and the trace-free Ricci tensor $S_\mn$ are (see, e.g., \cite{gst1})
\begin{eqnarray}
&&g^{\m\n}=\bar{g}^{\m\n}-2Vl^\m l^\n,\label{inv}\\
&&\nn\\
&&G_\mn=-\frac{(D-1)(D-2)}{2}Kg_\mn+S_\mn,~~~S_{\mu \nu}=-\rho l_{\mu}\, l_{\nu},\label{GKS}
\end{eqnarray}
where
\begin{equation}\label{rho}
  \rho=\l[\bar{\Box}+2\xi^\a\pa_\a +\frac{1}{2}\xi_\a\xi^\a+2(D-2)K\r]V\equiv -\mathcal{O} V,
\end{equation}
with $\bar{\Box}\equiv\bar{\nabla}_\m\bar{\nabla}^\m$ and $\bar{\nabla}_\m$ being the covariant derivative with respect to the background metric $\bar{g}_{\m\n}$.

In this work, we consider the most general gravity theory coupled with electromagnetic field. The Lagrange function of the whole theory depends on the curvature tensor, the electromagnetic field, and their covariant derivatives at any order. We call such a theory Generic Einstein-Maxwell Theory. We then assume that the spacetime metric is of the KSK form defined above. With this assumption, we prove a theorem stating that the KSK metrics simplify the field equations of any generic Einstein-Maxwell theory. To prove this theorem, we use the technique that has been used in Ref.\cite{gst4}. As an explicit example, we examine the Horndeski's vector-tensor theory \cite{horn}, which generalises the Einstein-Maxwell theory by adding some special curvature-electromagnetic couplings, and write its field equations in the KSK spacetimes.

Our paper is structured as follows. In Sec. 2, we review the universality of the KSK metrics for a generic gravity theory. In Sec. 3, we give the generalization of the universality property given in Sec. 2 by considering a generic gravity theory with Maxwell's field. In Sec. 4, we present the Horndeski's vector-tensor theory as an explicit example for our formulation, and we conclude in Sec. 5.

\section{Universality of Kerr-Schild-Kundt Metrics}

In a recent paper  \cite{gst4}, it has been proved that the KSK metrics given by the form (\ref{KS}) satisfying (\ref{lxi}) and (\ref{lV}) simplifies the field equations of any generic gravity theory constructed from the Riemann tensor and its covariant derivatives at any orders. Here we shall now give a brief review of this property--\textit{universality}--of the KSK metrics.

A vacuum generic gravity theory can be described by the action
\begin{equation}\label{gG}
  I=\int d^Dx \sqrt{-g}f(g,R,\nabla R,\ldots),
\end{equation}
where $f$ is a smooth function of  the metric tensor $g$, the Riemann tensor $R$, the covariant derivative of Riemann tensor $\nabla R$, and the higher order covariant derivatives of $R$, respectively. For the KSK metrics, it can be shown that the field equations of the theory (\ref{gG}), obtained by variation with respect to the metric $g_\mn$, take the form (see, e.g., \cite{gst4})
\begin{equation}
E_{\mu\nu}\equiv eg_{\mu\nu}+\sum_{n=0}^{N}\,a_{n}\square^{n} S_{\mu\nu}=0, \label{gGEqns}
\end{equation}
where $S_\mn$ is the traceless Ricci tensor and $\Box$ is the d'Alembertian with respect to $g_\mn$. The derivative order of the theory becomes $2N+2$ such that $N=0$ represents the Einstein's gravity and $N=1$ represents the quadratic curvature gravity or more generally F(Riemann) theories. Taking the trace of (\ref{gGEqns}) produces the scalar equation
\begin{equation}\label{eEqn}
  e=0,
\end{equation}
which determines the effective cosmological constant in terms of the parameters of the theory. Inserting (\ref{eEqn}) into (\ref{gGEqns}) produces the traceless part
\begin{equation}
\sum_{n=0}^{N}\,a_{n}\square^{n} S_{\mu\nu}=0, \label{SEqns}
\end{equation}
which must be satisfied independently. This is a nontrivial nonlinear differential equation which cannot be solved in general, except for some trivial cases. However, it has been shown in \cite{gst2} that (\ref{SEqns}) can also be written as the linear equation
\begin{equation}
l_\m l_\n\sum_{n=0}^{N}\,a_{n}(-1)^n\l(\mathcal{O}-2K\r)^{n}\mathcal{O}V=0, \label{SEqns1}
\end{equation}
since $S_\mn=-\rho l_\m l_\n$ and
\begin{equation}
\square^{n}S_{\mu\nu}=(-1)^n l_\m l_\n\l(\mathcal{O}-2K\r)^{n}\mathcal{O}V, \label{}
\end{equation}
for the KSK metrics. Here, $\mathcal{O}$ is the operator defined in (\ref{rho}). This result is true for any $\xi_\m$ satisfying $l_\m \xi^\m=0$, the first condition in (\ref{lV}). For $N\geq1$, it is further possible to factorize (\ref{SEqns1}) as
\begin{equation}
\prod_{n=0}^{N}(\mathcal{O}+b_n)\mathcal{O}V=0, \label{SEqns2}
\end{equation}
where $b_n$'s are related to $a_n$'s and so to the parameters of the theory. Now if all $b_n$'s are distinct and nonzero, the most general solution of (\ref{SEqns2}) can be given in the form
\begin{equation}\label{}
  V=V_E+V_1+V_2+\ldots+V_N,
\end{equation}
where $V_E$ is the solution of the Einstein gravity equation
\begin{equation}\label{EinEqn}
  \mathcal{O}V_E=0,
\end{equation}
and each $V_n$, for $n=1,2,\ldots,N$, is the solution of the quadratic curvature gravity equation
\begin{equation}\label{QuadEqn}
  (\mathcal{O}+b_n)V_n=0.
\end{equation}
At this point, it is worth mentioning that there are some special cases in which some or all of $b_n$'s coincide or vanish. In these cases, fourth or higher power operators, such as $(\mathcal{O}+b_n)^2$, appear and Log-type solutions, which exist in the so-called critical theories, arise in the solution spectrum of the generic gravity theory. The equations (\ref{EinEqn}) and (\ref{QuadEqn}) can easily be solved for $V_E$ and $V_n$ by using such techniques as the method of separation of variables and the method of Green's function.

In the proof of the universality theorems for the KSK metrics \cite{gst4}, we use some properties of the null vector $l_{\mu}$. First, note that the contractions of $l^{\mu}$ with
$l_{\mu}$, $\xi_{\mu}$, and $\partial_{\mu}V$ yield zero. Secondly, the contractions of $l^{\mu}$ with the first order derivatives of $\xi_{\mu}$ and $\partial_{\mu}V$ yield
\begin{eqnarray}
&&l^{\nu}\nabla_{\mu}\xi_{\nu}=-\frac{1}{2}l_{\mu}\xi^{\nu}\xi_{\nu}, \label{eq:1st_id}\\
&&\nabla_{\mu}\xi^{\mu}=-\frac{1}{4}\xi^{\mu}\xi_{\mu}+\frac{2D-3}{D\left(D-1\right)}R, \label{eq:Div_of_ksi}\\
&&l^{\mu}\nabla_{\mu}\xi_{\alpha}=-l_{\alpha}\left(\frac{1}{4}\xi^{\mu}\xi_{\mu}-\frac{1}{D\left(D-1\right)}R\right),\label{eq:Directional_deriv_of_ksi}\\
&&l^{\mu}\nabla_{\mu}\partial_{\nu}V=l^{\mu}\nabla_{\nu}\partial_{\mu}V=-\frac{1}{2}l_{\nu}\xi^{\mu}\partial_{\mu}V.\label{eq:lambda_cont_with_1st_order_derv_of_V}
\end{eqnarray}
So, here are the important points to observe (see \cite{gst4} for more details):
\begin{itemize}
  \item The number of  $l$ vectors is preserved since a free-index $l$ always appears in the results;
  \item The contraction with the $l$ vector removes the first order derivatives acting on $\xi_{\mu}$ and $\partial_{\mu}V$;
  \item The contraction of the $l$ vector with the higher order derivatives of $\xi_{\mu}$ and $\partial_{\mu} V$ produce free-indexed $l$ vectors.
\end{itemize}
We define the $l$-degree of a tensor as the number of free-indexed $l$ vectors contained. For example, the $l$-degree of the Weyl tensor is two \cite{gst4}. According to this definition, from the above discussions, we can say that the contraction of the $l^\m$ vector with the covariant derivatives of the vectors $\xi_{\mu}$ and $\partial_{\mu} V$ preserves the $l$-degree of the relevant tensor. Our definition of $l$-degree of a tensor is equivalent to boost-weight of a tensor defined by Coley et al. \cite{hervik1} (and see the references therein).

\section{Generic Gravity Theories with Maxwell's Field}

Now we wish to extend the theorem given in Sec. 2 on the universality of the KSK metrics \cite{gst4} to generic gravity theories with the Maxwell field. The Lagrange function of such a theory should contain the metric tensor $g_{\mu \nu}$ and its inverse $g^{\mu \nu}$, the Riemann tensor $R_{\alpha \beta \mu \nu}$, the Maxwell's field tensor $F_{\mu \nu}$, and the covariant derivatives of these tensors of all orders. That is, in $D$ dimensions, the most general action for the Einstein-Maxwell theory is
\begin{equation}\label{actgen}
I=\int d^Dx \sqrt{-g}\,L(g,R,\nabla\nabla\ldots\nabla R,F,\nabla\nabla\ldots\nabla F),
\end{equation}
Let the the electromagnetic vector potential be given by $A_{\mu}= \phi\,l_{\mu}$ where $\phi$ is a function satisfying the condition $l^{\mu}\, \phi_{,\mu}=0$.
Then, the Maxwell field tensor takes the form
\begin{equation}\label{}
F_{\mu \nu}=\nabla_\m A_\n-\nabla_\n A_\m=\phi_{,\mu} l_{\nu}-\phi_{,\nu} l_{\mu},
\end{equation}
which satisfies the following conditions
\begin{eqnarray}
F_\mn F^\mn=0,\\
l_{\mu}\, F^{\mu \nu}=0, \\
F_{\mu \alpha}\, F_{\nu}\, ^{\alpha}=\psi l_{\mu}\, l_{\nu},
\end{eqnarray}
where $\psi=g^{\mu \nu}\, \phi_{,\mu}\, \phi_{,\nu}$. For the extension of the universality theorem to generic gravity theories with an antisymmetric tensor $F_{\alpha \beta}$, we use the following notation:

 \begin{itemize}

 \item $\nabla^n F$ denotes $n$-number of covariant derivatives of the $F$ tensor, ie., $\nabla_{\alpha_{1}}\,\nabla_{\alpha_{2}} \cdots \nabla_{\alpha_{n}}\, F_{\mu \nu}$.

 \item $[(\nabla^{n} F)\, (\nabla ^{m} F)]_{\mu \nu}$ denotes a second rank symmetric tensor obtained form the product tensors $(\nabla^{n} F)\, (\nabla ^{m} F)$ of rank $(4+m+n)$.

 \end{itemize}

\noindent With all these, we now have the  following theorem:
\begin{quotation}
\noindent \textbf{Theorem 1:} \emph{Let the spacetime metric be given by the Kerr-Schild-Kundt (KSK) type
\[
g_{\mu\nu}=\bar{g}_{\mu\nu}+2Vl_{\mu}l_{\nu},
\]
with the properties
\[
l^{\mu}l_{\mu}=0,\qquad\nabla_{\mu}l_{\nu}=\xi_{(\mu}l_{\nu)},\qquad\xi_{\mu}l^{\mu}=0,\qquad l^{\mu}\partial_{\mu}V=0,
\]
and let the electromagnetic vector potential $A_{\mu}=\phi\,l_{\mu}$, or the Maxwell's field tensor
\[
F_{\mu \nu}=\phi_{,\mu} l_{\nu}-\phi_{,\nu} l_{\mu},
\]
with the property $l^{\mu} \phi_{,\mu}=0$, where $\bar{g}_{\mu\nu}$ is the metric of a space of constant curvature (A)dS. Then any second rank symmetric tensor constructed from the Riemann tensor, Maxwell's field tensor, and their covariant derivatives can be written as a linear combination of $g_{\mu\nu}$, $S_{\mu\nu}$, $F_{\mu}\,^{\alpha}\,F_{\nu \alpha}$, and their higher derivatives in the form $\square^{n}S_{\mu\nu}$ and $[(\nabla^{n} F)\, (\nabla ^{m} F)]_{\mu \nu}$ for all $m$ and $n$, where $\square$ represents the d'Alembertian with respect to $g_{\mu\nu}$; that is,
\begin{equation}
E_{\mu\nu}\equiv eg_{\mu\nu}+\sum_{n=0}^{N}\,a_{n}\square^{n}\, S_{\mu\nu}+\sum_{m=0, n=0}^{M}\, b_{mn}\,[(\nabla^{n} F)\, (\nabla ^{m} F)]_{\mu \nu}, \label{denk1}
\end{equation}
and
\begin{equation}
E^{\mu}\equiv \nabla_{\alpha}\,\left[\sum_{n=0}^{M}c_{n}\square^{n}\, F^{\alpha \mu}\right], \label{denk2}
\end{equation}
where  $a_{n}, b_{mn}$ and $c_{n}$ are constants coming from the parameters of the theory and $N$ and $M$ are numbers related to the derivative orders in the theory. Then the associated field equations of the generic Einstein-Maxwell theory are $E_{\mu\nu}=0$ and $E_{\mu}=0$.
}

\end{quotation}

\vspace{0.5cm}
\noindent{\bf Sketch of the Proof}:
\vspace{0.4cm}

The most general Lagrange function for the generic Einstein-Maxwell theory given in (\ref{actgen}) can be written as follows
\begin{eqnarray}
L&=&L_{1}(g,R, \nabla R, \nabla \nabla R, \ldots)+L_{2}(g,R, \nabla R, \nabla \nabla R, \ldots, F, \nabla \nabla F, \ldots)\nn\\
&&+L_{3}(g,F, \nabla F, \nabla \nabla F, \ldots),
\end{eqnarray}
where $L_{1}$ is a function of the curvature tensor and its covariant derivatives of any order, $L_{2}$ is a function representing the coupling of the electromagnetic tensor $F$ and the curvature tensor $R$ at any order, and $L_{3}$ is a function depending solely on $F$ and its covariant derivatives of any order. Then the field equations associated with the above Lagrange function can be written as
\begin{eqnarray}
&&E^{1}_{\mu \nu}+E^{2}_{\mu \nu}+E^{3}_{\mu \nu}=0, \\
&&E^{2}_{\mu}+E^{3}_{\mu}=0,
\end{eqnarray}
where $E^{1}_{\mu \nu}, E^{2}_{\mu \nu}, E^{3}_{\mu \nu}$ are the tensors obtained from the variation of the action (\ref{actgen}) with respect to the metric tensor and $E^{2}_{\mu}$ and $E^{3}_{\mu}$ are the vectors obtained from the variation of the action (\ref{actgen}) with respect to the electromagnetic vector potential vector $A_{\mu}$. All of these two-rank
symmetric tensors and the vectors have the following forms in general:
\begin{eqnarray}
&&E^{1}_{\mu \nu}=eg_\mn+\sum_{n_{0},n_{1}, \cdots , n_{k}}\,C^{1}_{n_{0},n_{1}, \dots n_{k}} \left[R^{n_{0}} \nabla^{n_{1}} R \nabla^{n_{2}} R \cdots \nabla^{n_{k}} R \right]_{\mu \nu},\label{E1mn}\\
&&E^{2}_{\mu \nu}= \nonumber \\
&&\sum_{n_{0},n_{1}, \cdots , n_{k},s_{0},s_{1}, \cdots s_{k}}\,C^{2}_{n_{0},n_{1}, \dots n_{k},s_{0}, s_{1}, \cdots s_{k}} \, \left[R^{n_{0}} \nabla^{n_{1}} R \nabla^{n_{2}} R \cdots \nabla^{n_k} R\, F^{s_{0}} \nabla^{s_{1}} F \cdots \right]_{\mu \nu},\label{E2mn}\\
&&E^{3}_{\mu \nu}=\sum_{t_{0}, t_{1}, \cdots t_{k}}\, C^{3}_{t_{0} t_{1} \cdots t_{k}} \left[F^{t_{0}} \nabla^{t_{1}} F \cdots \nabla^{t_{k}} F \ \right]_{\mu \nu},\label{E3mn}
\end{eqnarray}
where the coefficients $C^{1}$, $C^{2}$ and $C^{3}$ are all constants and $e$ is a function of scalars obtained from the Riemann tensor and the electromagnetic field tensor and their covariant derivatives, and
\begin{eqnarray}
&&E^{2}_{\mu}= \nonumber \\
&&\sum_{n_{0},n_{1}, \cdots , n_{k},s_{0},s_{1}, \cdots s_{k}}\,C^{4}_{n_{0},n_{1}, \dots n_{k},s_{0}, s_{1} \cdots s_{k}} \, \left[R^{n_{0}} \nabla^{n_{1}} R \nabla^{n_{2}} R \cdots \nabla^{n_k} R F^{s_{0}} \nabla^{s_{1}} F \cdots \right]_{\mu},\label{E2m}\\
&&E^{3}_{\mu}=\sum_{t_{0}, t_{1}, \cdots t_{k}}\, C^{5}_{t_{0} t_{1} \cdots t_{k}} \left[F^{t_{0}} \nabla^{t_{1}} F \cdots \nabla^{t_{k}} F \ \right]_{\mu},\label{E3m}
\end{eqnarray}
where $C^{4}$ and $C^{5}$ are constants. To proceed further, we now consider typical monomials in each of $E^{1}_{\mu \nu}, E^{2}_{\mu \nu}, E^{3}_{\mu \nu}$, and $E^{2}_{\mu}$ and $E^{3}_{\mu}$.

After inserting the KSK metric tensor into (\ref{E1mn}) and using $R_{\mu \nu \alpha \beta}=K (g_{\mu \alpha} g_{\nu \beta}-g_{\nu \alpha} g_{\mu \beta})+r_{\mu \nu \alpha \beta}$, where $r_{\mu \nu \alpha \beta}$ is a tensor depending on the vectors $\xi_{\mu}$, $\partial_{\nu} V$, and their covariant derivative at any order (see \cite{gst4} for the explicit expression), one can reduce $E^{1}_\mn$ to
\begin{equation}
E^{1}_{\mu \nu}= e_{0} g_{\mu \nu}+\sum_{n_{0},n_{1}, \cdots , n_{k}}\,\bar{C}^{1}_{n_{0},n_{1}, \dots n_{k}} \left[r^{n_{0}} \nabla^{n_{1}} r \nabla^{n_{2}} r \cdots \nabla^{n_{k}} r \right]_{\mu \nu},
\end{equation}
where $e_{0}$ is a constant and $\bar{C}^{1}$ are constants. A typical monomial in $E^{1}_\mn$ is, therefore,
\begin{equation}
\left[r^{n_{0}} \nabla^{n_{1}} r \nabla^{n_{2}} r \cdots \nabla^{n_{k}} r \right]_{\mu \nu}. \label{monr}
\end{equation}
Since the $l$-degree of $r_{\mu \nu \alpha \beta}$ is two, the number of free $l$ vectors in such a monomial is $2 n_{0}+2k$. Since the contraction of
the l vector with $\xi_{\mu}$ and $\partial_{\mu} V$ yields zero, and with their
covariant derivatives of any order, this keeps the number of
free l-vectors unchanged, in order to have a nonzero term in
the monomial [Eq. (\ref{monr})] at the end of the contractions, it
must be that $2n_{0}+2 k=2$, which can only be satisfied
either when $n_{0}=1$ or when $k=1$.
This means that \cite{gst4}
\begin{equation}
E^{1}_{\mu \nu}=e_{0} g_{\mu \nu}+\rho_{1} S_{\mu \nu}+\rho_{2} [\nabla \nabla \cdots \nabla r]_{\mu \nu}, \label{E11mn}
\end{equation}
where $\rho_1$ and $\rho_2$ are some scalars containing $V$, $\p$, and their partial derivatives. This result is equivalent to (by the use of Bianchi identities) \cite{gst4}
\begin{equation}
E^{1}_{\mu\nu}=e g_{\mu\nu}+\sum_{n=0}^{N}\,a_{n}\square^{n}\, S_{\mu\nu}.
\end{equation}
A typical monomial of $E^{2}_\mn$ can be written from (\ref{E2mn}) as
\begin{equation}
\left[R^{n_{0}} \nabla^{n_{1}} R \nabla^{n_{2}} R \cdots \nabla^{n_k} R\, F^{s_{0}} \nabla^{s_{1}} F \cdots \nabla^{s_{k}} F \right]_{\mu \nu}.
\end{equation}
When we insert $R_{\mu \nu \alpha \beta}=K (g_{\mu \alpha} g_{\nu \beta}-g_{\nu \alpha} g_{\mu \beta})+r_{\mu \nu \alpha \beta}$ in the above monomial, the terms coming from the $K$ part of the curvature tensor reduce to monomials with less number of $r$ tensors and also to the  monomials containing only Maxwell fields; i.e., they join to $E^{3}_{\mu \nu}$. There will be no contributions to the $e_{0}$ part of the field equations $E^{1}_{\mu \nu}$ in (\ref{E11mn}) from such monomials. The remaining part of the monomial will therefore be exactly of the above form but instead of $R$'s we have now $r$'s:
\begin{equation}
\left[r^{n_{0}} \nabla^{n_{1}} r \nabla^{n_{2}} r \cdots \nabla^{n_k} r\, F^{s_{0}} \nabla^{s_{1}} F \cdots \nabla^{s_{k}} F \right]_{\mu \nu}.
\end{equation}
For the KSK ansatz, we let $A_{\mu} =\phi \,l_{\mu}$, where $\phi$ is a function satisfying $l^{\mu} \partial _{\mu} \phi=0$, and $F_{\mu \nu}=\phi_{,\mu} l_{\nu}-\phi_{,\nu} l_{\mu}$. Then, the $l$-degrees of $r_{\mu \nu \alpha \beta}$ and $F_{\mu \nu}$ are two and one respectively. The number of free $l$-vectors in the bracket is $2n_{0}+2k+s_{0}+k$ which must be equal to two for having non-vanishing terms. Since these monomials must contain both $r$ and $F$ tensors, then
it is easy to see that $2n_{0}+2k+s_{0}+k>2$; for this reason, all such coupling terms must vanish. This means that, for KSK metrics and for $F_{\mu \nu}=\phi_{,\mu} l_{\nu}-\phi_{,\nu} l_{\mu}$, there will be no coupling of the tensors $r$ and $F$; such terms vanish identically. A typical monomial of $E^{3}_\mn$ in (\ref{E3mn}) can be given as
\begin{equation}
\left[F^{t_{0}} \nabla^{t_{1}} F \cdots \nabla^{t_{k}} F \ \right]_{\mu \nu}.
\end{equation}
The number of free $l$-vectors in this expression is $t_{0}+k$. After contractions, this number will be preserved and hence, for non-vanishing terms, we must have $t_{0}+k=2$. This means that either $t_{0}=2$ ($F^2$ term) or $k=2$ ($\nabla \nabla \cdots \nabla F \nabla \nabla \cdots \nabla F$ terms) or $t_{0}=1, k=1$ (symmetrized $F \nabla \nabla \cdots \nabla F$ terms). Combining these, we get
\begin{equation}
E^{3}_{\mu \nu}=\sum_{m=0, n=0}^{M}\, b_{mn}\,[(\nabla^{n} F)\, (\nabla ^{m} F)]_{\mu \nu},
\end{equation}
where $b_{mn}$'s are constants. This completes the proof of the first part of Theorem 1.

To prove the second part of the theorem we use the same approach. A typical monomial of $E^2_{\mu}$ in (\ref{E2m}) can be written as
\begin{equation}
\left[R^{n_{0}} \nabla^{n_{1}} R \nabla^{n_{2}} R \cdots \nabla^{n_k} R F^{s_{0}} \nabla^{s_{1}} F \cdots \nabla^{s_{k}} F \right]_{\mu}
\end{equation}
After inserting $R_{\mu \nu \alpha \beta}=K (g_{\mu \alpha} g_{\nu \beta}-g_{\nu \alpha} g_{\mu \beta})+r_{\mu \nu \alpha \beta}$, the terms related to the $K$ part of the curvature tensors in the above monomials reduce either to the same type of monomials with less number of $r$'s or to monomials of $E^{3}_{\mu}$ in (\ref{E3m}). Hence, we can study the above monomials only with $r$'s instead of $R$'s. In such a case, the number of free $l$-vectors is $2n_{0}+2k+s_{0}+k$. To have non-zero terms in the monomial, we must have $2n_{0}+2k+s_{0}+k=1$, but this is not possible because such monomials represent couplings between $r$ and $F$ tensors and so $2n_{0}+2k+s_{0}+k\neq1$ for all cases. That is to say, $E^{2}_{\mu}=0$ identically. Finally, a typical monomial of $E^3_{\mu}$ in (\ref{E3m}) can be given by
\begin{equation}
\left[F^{t_{0}} \nabla^{t_{1}} F \cdots \nabla^{t_{k}} F \ \right]_{\mu}.
\end{equation}
The number of free $l$-vectors in the above expression is $t_{0}+k$. For non-zero terms, this number must be equal to one; therefore, we must have either $t_{0}=1$ (not possible) or $k=1$ ($\nabla \nabla \cdots \nabla F$ terms). Thus, we obtain, by the use of Bianchi identities,
\begin{equation}
E^3_{\mu}= \nabla_{\alpha}\,\left[\sum_{n=0}^{M}c_{n}\square^{n}\, F^{\alpha}\,_{ \mu}\right]=0.\label{denk2-2}
\end{equation}
This completes the proof of Theorem 1.

\vspace{0.3cm}

\noindent
{\bf Remark 1}: In the case of the KSK metrics, it is straightforward to show that the Maxwell equations in (\ref{denk2}) and (\ref{denk2-2}) can also be written as
\begin{equation}
E^{\mu}\equiv \sum_{n=0}^{M}\bar{c}_{n}\square^{n}(\nabla_{\alpha} F^{\alpha \mu})=0.
\end{equation}

\vspace{0.3 cm}

For the KSK metrics, the trace of (\ref{denk1}) reduces to $e=0$, which gives a relation between the parameters of the theory and the cosmological constant, and the remaining part of (\ref{denk1}) gives
\begin{equation}
\sum_{n=0}^{N}a_n\,{\cal O}^{n+1}V+\rho_e=0, \label{denk3}
\end{equation}
where $\rho_e$ is the source term for the equation of $V$ and the operator ${\cal O}$ is defined in (\ref{rho}), namely it is given by
\begin{equation}\label{opq}
{\cal O} V=-\l[ \bar{\Box}+2\xi^\a\pa_\a +\frac{1}{2}\xi_\a\xi^\a+2(D-2)K \r]V.
\end{equation}
On the other hand, the equation (\ref{denk2}) reduces to
\begin{equation}
\sum_{n=0}^{M}\,c_{n}{\cal R}^n \eta=0, \label{denk3-1}
\end{equation}
where
\begin{equation}
\eta=\bar{\square} \phi+\xi^{\alpha}\, \phi_{,\alpha},
\end{equation}
and the operator ${\cal R}$ is defined by
\begin{equation}
{\cal R} \eta=[\bar{\square}+\xi^{\alpha}\, \partial_{\alpha} +(D-1)\,K]\eta.
\end{equation}
To derive the above operators, we used the following identities:
\begin{eqnarray}
&&\square\, l_{\mu}=(D-1)\,K\,l_{\mu},  \\
&&l^{\alpha}\, \bar{\nabla}_{\alpha}\,
\xi_{\mu}=\l(K-\frac{1}{4}\, \xi^{\alpha}\,
\xi_{\alpha}\r) l_{\mu},  \\
&&\bar{\nabla}^{\alpha}\, \xi_{\alpha}+\frac{1}{4}\,\xi^{\alpha}\,
\xi_{\alpha}-(2D-3)\,K=0.
\end{eqnarray}
and
\begin{equation}\label{denk7}
\bar{\nabla}_{\nu}\, \xi_{\beta}=\frac{1}{2}\, \xi_{\nu}\,
\xi_{\beta}+2K\, \bar{g}_{\nu \beta}+n_{\nu}\,
l_{\beta}+2 n_{\beta}\, l_{\nu}-\mu l_{\nu}\,
l_{\beta},
\end{equation}
where $\mu$ is a function and $n_{\mu}$ is a vector satisfying
\[
l^{\alpha}\, n_{\alpha}=-\frac{1}{4}\, \xi^2-K,
\]
where $\xi^2=\xi^{\alpha}\, \xi_{\alpha}$. We also have
\[
\nabla_{\nu}\xi^2=(\xi^2+4K)\, \xi_{\nu}+4\, (\xi^{\alpha}\,
n_{\alpha})\, l_{\nu}.
\]
Since $p_{\mu}\, l^{\mu}=0$, where $p_\m\equiv\pa_\m\p$, it is now easy to calculate
\begin{equation}\label{xipp}
(\bar{\nabla}_{\mu} \xi_{\nu}\,) p^{\mu} p^{\nu}=\frac{1}{2}\, (\xi_{\mu}\,p^{\mu})^2+2K\, p^{\mu}\,p_{\mu}.
\end{equation}

As a final remark, by using the three steps below, we can express $[(\nabla^{n} F)\, (\nabla ^{m} F)]_{\mu \nu}$ in (27) for any $n$ and $m$ proportional to $l_\m l_\n$:
\begin{itemize}
  \item The number of  $l$ vectors is preserved since a free-index $l$ always appears in the results;
  \item The contraction with the $l$ vector removes the first order derivatives acting on $\xi_{\mu}$, $\partial_{\mu}V$, and $\partial_{\mu}\p$; and
  \item The contraction of the $l$ vector with the higher order derivatives of $\xi_{\mu}$, $\partial_{\mu}V$, and $\partial_{\mu} \p$ produce free-indexed $l$ vectors.
\end{itemize}
For illustration, we give the special cases (i) Lagrange function depends only on $F$ and (ii) Lagrange function depends on $F$ up to the first order covariant derivatives as the following corollaries.

\vspace{0.5cm}

\noindent
{\bf Corollary  1}: If Lagrange function contains only $F$'s (no derivatives) then the reduced field equations are

\begin{equation}
E_{\mu\nu}=eg_{\mu\nu}+\sum_{n=0}^{N} \,a_{n}\square^{n}\, S_{\mu\nu}+b \tau_{\mu\nu}=0, \label{denk3-2}
\end{equation}
and
\begin{equation}
E^{\mu}= \nabla_{\alpha}\, F^{\alpha \mu}=0, \label{denk4}
\end{equation}
where $\tau_{\mu \nu}=F_{\mu}\,^{\alpha}\,F_{\nu \alpha}=\psi l_{\mu} l_{\nu}$.

\vspace{0.5cm}

\noindent
{\bf Corollary 2}: If Lagrange function contains $F$'s and first derivatives of $F$'s then the reduced field equations are

\begin{eqnarray}
&&E_{\mu\nu}=eg_{\mu\nu}+\sum_{n=0}^{N} \,a_{n}\square^{n}\, S_{\mu\nu}+b_{1} \tau_{\mu\nu}+b_{2}\, \square \tau_{\mu \nu}+b_{3}\, \nabla_{\gamma} F_{\mu \alpha}\, \nabla^{\gamma} F_{\nu}\,^{\alpha} \nonumber \\
&&~~~~~~~~+b_{4}\,\nabla_{\mu} F_{\beta \alpha}\, \nabla_{\nu} F^{\beta}\,^{\alpha}+b_{5}\,\nabla_{\alpha} F_{\beta \mu}\, \nabla^{\beta} F^{\alpha}\,_{\nu}=0, \label{denk4-1}
\end{eqnarray}
and
\begin{equation}
E^{\mu}= c_{1} \nabla_{\alpha}\, F^{\alpha \mu}+c_{2}\,\nabla_{\alpha}\square F^{\alpha \mu}=0. \label{denk5}
\end{equation}

\vspace{0.5cm}

\noindent
{\bf Remark 2}: In Corollary 2, there are five different symmetric tensors obtained by the first derivatives of $F$'s:
\begin{eqnarray}
&&1)~~~ (\square F_{\mu \alpha})\, F_{\nu}\,^{\alpha}+F_{\mu \alpha} \square F_{\nu}\,^{\alpha} ,\\
&&2)~~~ \square (F_{\mu \alpha} F_{\nu}\, ^{\alpha}), \\
&&3)~~~ \nabla_{\gamma} F_{\mu \alpha}\, \nabla^{\gamma} F_{\nu}\,^{\alpha}, \\
&&4)~~~ \nabla_{\mu} F_{\beta \alpha}\, \nabla_{\nu} F^{\beta}\,^{\alpha}, \\
&&5)~~~ \nabla_{\alpha} F_{\beta \mu}\, \nabla^{\beta} F^{\alpha}\,_{\nu}.
\end{eqnarray}
But the first and second terms are not independent, can be expressed in terms of the others:
\[
\square (F_{\mu \alpha}\,F_{\nu}\,^{\alpha})=2\nabla_{\gamma} F_{\mu \alpha}\, \nabla^{\gamma} F_{\nu}\,^{\alpha}+(\square F_{\mu \alpha})\, F_{\nu}\,^{\alpha}+F_{\mu \alpha} \square F_{\nu}\,^{\alpha}.
\]
Using the Bianchi identity for $F$'s, we get
\[
\nabla_{\mu} F_{\beta \alpha}\, \nabla_{\nu} F^{\beta}\,^{\alpha}=2 \nabla_{\gamma} F_{\mu \alpha}\, \nabla^{\gamma} F_{\nu}\,^{\alpha}-2 \nabla_{\alpha} F_{\beta \mu}\, \nabla^{\beta} F^{\alpha}\,_{\nu}.
\]
On the other hand, for the KSK metric and $M=2$, we find
\begin{eqnarray}
&& \nabla_{\gamma} F_{\mu \alpha}\, \nabla^{\gamma} F_{\nu}\,^{\alpha}=\rho_{1} l_{\mu} l_{\nu}, \\
&&\nabla_{\mu} F_{\beta \alpha}\, \nabla_{\nu} F^{\beta}\,^{\alpha}=\rho_{2} l_{\mu} l_{\nu},\\
&&\nabla_{\alpha} F_{\beta \mu}\, \nabla^{\beta} F^{\alpha}\,_{\nu}=\rho_{3} l_{\mu} l_{\nu},
\end{eqnarray}
where
\begin{eqnarray}
\rho_{1}&=&-\frac{1}{2} (p_{\alpha} \xi^{\alpha})^2+(\nabla_\a p_\b+\frac{1}{2} p_{\alpha} \xi_{\beta})(\nabla^\alpha p^\beta
+\frac{1}{2} p^{\alpha} \xi^{\beta}), \\
\rho_{2}&=&-(p_{\alpha} \xi^{\alpha})^2+\frac{1}{2} (p_{\alpha}\, p^{\alpha})(\xi^{\beta}\, \xi_{\beta}),\\
\rho_{3}&=&\nabla_\a p_\b\nabla^\alpha p^\beta+p^{\alpha} \xi^{\beta}\nabla_\a p_\b,
\end{eqnarray}
so that $\rho_{2}=2(\rho_{1}-\rho_{3})$ and hence we can set $b_{5}=0$. Here we defined $p_\m\equiv\pa_\m\p$.

\section{Horndeski's Vector-Tensor Theory: An Explicit Example}

As an explicit example, we shall consider the Horndeski's Vector-Tensor Theory which is a generalization of Einstein-Maxwell theory that
leads to second order equations of motion and satisfying charge conservation. This theory is described, in $D$ dimensions, by the action \cite{horn}
\begin{equation} 
I=\int d^{D}x\sqrt{-g}\,\l[\frac{R-2\La}{2\k^2}-\frac{1}{4}F_\mn F^\mn+\sigma\, \mathcal{R}_{~~\ab}^\mn F_\mn F^\ab \r],\label{action}
\end{equation}
where the parameters $\k^2$, $\La$, and $\sigma$ are the gravitational constant, the cosmological constant, and the Horndeski coupling constant, respectively, and
\begin{eqnarray}
&&F_\mn\equiv \nabla_\m A_\n-\nabla_\n A_\m,\label{F}\\
&&\mathcal{R}_{~~\ab}^\mn\equiv-\frac{1}{4}\d^{\m\n\la\s}_{\a\b\rho\tau}R_{~~\la\s}^{\rho\tau}.\label{cR}
\end{eqnarray}
Making explicit use of the generalized Kronecker delta defined by
\begin{equation}\label{}
  \d^{\a_1\ldots\a_k}_{\b_1\ldots\b_k}=k!\d^{[\a_1}_{\b_1}\ldots\d^{\a_k]}_{\b_k}=k!\d^{\a_1}_{[\b_1}\ldots\d^{\a_k}_{\b_k]},
\end{equation}
one can show that the Horndeski interaction term in (\ref{action}) can also be written as
\begin{equation}\label{}
  \mathcal{R}_{~~\ab}^\mn F_\mn F^\ab=-RF^2+4R_\m^{~\n} F_{\n\a}F^{\m\a}-R^\mn_{~~\ab} F_\mn F^\ab,
\end{equation}
where $F^2\equiv F_\mn F^\mn$ and $R_\m^{~\n}$ is the Ricci tensor. The field equations derived from the action (\ref{action}) are
\begin{eqnarray}
&&G_\m^{~\n}+\La \d_\m^\n=\k^2(T_\m^{~\n}+\sigma\,\tau_\m^{~\n}),\label{eqn01}\\
&&\nabla_\n \mathcal{F}^\mn=0,\label{eqn02}
\end{eqnarray}
where
\begin{eqnarray}
&&T_\m^{~\n}\equiv F_{\m\a}F^{\n\a}-\frac{1}{4}\d_\m^\n F^2,\label{TEM}\\
&&\tau_\m^{~\n}\equiv\d^{\n\a\b\g}_{\m\rho\s\tau}\nabla_\a F^{\s\tau}\nabla^\rho F_{\b\g}-4\mathcal{R}_{~~\m\a}^{\n\rho} F_{\rho\k} F^{\a\k},\label{TH}\\
&&\mathcal{F}^\mn \equiv F^\mn-4\sigma\mathcal{R}_{~~\ab}^\mn F^\ab.   \label{cF}
\end{eqnarray}


Now using the KSK ansatz (\ref{KS}) having the properties (\ref{Rback})-(\ref{GKS}) together with the electromagnetic vector potential of the form
\begin{equation}\label{pot}
  A_\m=\p(x)l_\m,
\end{equation}
where $l_\m$ and $p_\m\equiv\pa_\m\p$ satisfies $l_\m p^\m=0$, one can show that
\begin{eqnarray}
&&F_\mn=p_{\m}l_{\n}-p_{\n}l_{\m},~~~~T_\m^{~\n}=\psi l_\m l^\n,\\
&&\tau_\m^{~\n}=-4\bl\{\bar{\nabla}_\alpha p_\b \bar{\nabla}^\a p^\b+\frac{1}{2}\xi^\a\pa_\a \psi-(\xi_\a p^\a)^2+\frac{1}{2}\l[\xi^2+(D-2)(D-3)K\r]\psi\br\}l_\m
l^\n,\label{THKS}\\
&&\mathcal{F}^\mn\equiv[1+4\s(D-2)(D-3)K]F^\mn,
\end{eqnarray}
where $\psi\equiv p_\m p^\m$. Then (\ref{eqn01}) and (\ref{eqn02}) become
\begin{eqnarray}
&&\l[\La-\frac{(D-1)(D-2)}{2}K\r]\d_\m^\n-\rho\, l_\m l^\n\nn\\
&&~~~~~~~~=\k^2\bl\{\psi-4\s\bl[\bar{\nabla}_\a p_\b \bar{\nabla}^\a p^\b+\frac{1}{2}\xi^\a\pa_\a \psi-(\xi_\a p^\a)^2\nn\\
&&~~~~~~~~~~~~~~~~~~~~~~~~~~~~~~~~~~~~~~~~+\frac{1}{2}\l[\xi^2+(D-2)(D-3)K\r]\psi\br]\br\}l_\m l^\n,\label{EinKS}\\
&&-[1+4\s(D-2)(D-3)K][\bar{\Box}\p+\xi^\n p_\n]l^\m=0.\label{delcF}
\end{eqnarray}
From these, we find that
\begin{eqnarray}
&&\La=\frac{(D-1)(D-2)}{2}K,\label{Lambda}\\
&&\nn\\
&&\bar{\Box}V+2\xi^\a\pa_\a V+\l[\frac{1}{2}\xi_\a\xi^\a+2(D-2)K\r]V\nn\\
&&~~~~=-\k^2\bl\{\psi-4\s\bl[\bar{\nabla}_\a p_\b \bar{\nabla}^\a p^\b+\frac{1}{2}\xi^\a\pa_\a \psi-(\xi_\a p^\a)^2\nn\\
&&~~~~~~~~~~~~~~~~~~~~~~~~~~~~~~~~~~~~~~+\frac{1}{2}\l[\xi^2+(D-2)(D-3)K\r]\psi\br]\br\}~,\label{EinKS1}\\
&&\bar{\Box}\p+\xi^\n p_\n=0.\label{Boxphi}
\end{eqnarray}
Observe that, in writing the last equation, we assumed the coefficient in (\ref{delcF}) is nonzero; i.e.,
\begin{equation}\label{sK}
  1+4\s(D-2)(D-3)K\neq0.
\end{equation}
Using the relation
\begin{equation}\label{}
  \bar{\nabla}_a p_\b \bar{\nabla}^\a p^\b=\frac{1}{2}\bar{\Box}\psi+\frac{1}{2}\xi^\a\pa_\a \psi-(D-1)K\psi+p^\a
  p^\b\bar{\nabla}_\a\xi_\b-p^\b\bar{\nabla}_\b(\bar{\Box}\p+\xi^\n\pa_\n\p),
\end{equation}
together with (\ref{xipp}) and (\ref{Boxphi}), we can equivalently write (\ref{EinKS1}) as
\begin{eqnarray}
&&\bar{\Box}V+2\xi^\a\pa_\a V+\l[\frac{1}{2}\xi_\a\xi^\a+2(D-2)K\r]V\nn\\
&&~~~~=-\k^2\bl\{\psi-4\s\bl[\frac{1}{2}\bar{\Box}\psi+\xi^\a\pa_\a \psi-\frac{1}{2}(\xi_\a p^\a)^2\nn\\
&&~~~~~~~~~~~~~~~~~~~~~~~~~~~~~~~~~~~~~~+\frac{1}{2}\l[\xi^2+(D-3)(D-4)K\r]\psi\br]\br\},\label{EinKS2}
\end{eqnarray}
Note that when $\xi_\m=0$ and $K=0$, all these expressions recover the flat background ($pp$-wave) case in Horndeski theory \cite{gh}. In a recent paper \cite{ghs}, we studied a modified version of this theory by adding extra couplings to (\ref{action}) of the form $R_{~~\ab}^\mn F_\mn F^\ab$ and obtained exact plane wave solutions to its field equations.

\vspace{0.3cm}

\noindent
{\bf Remark 3}: The equations (\ref{EinKS1}) and (\ref{Boxphi}) are special cases of the general field equations (\ref{denk3}) and (\ref{denk3-1}) for $n=0$. Furthermore, the Horndeski theory is a special case of Corollary 2 with no derivatives of $F_\mn$.

\vspace{0.3 cm}

\section{Conclusion}

In this work, we considered the most general Einstein-Maxwell theory in which the pure gravity and Maxwell parts and their couplings are thought to be
arbitrary. The Lagrange function associated with such a theory is any function of the curvature tensor, the electromagnetic field, and their covariant derivatives of any order. When the metric of the spacetime is assumed to be the Kerr-Schild-Kundt type of metrics, we proved a theorem stating that the most general Einstein-Maxwell field equations reduce to two coupled  simple equations for functions $V$ and $\phi$ representing the gravitational and electromagnetic potentials, respectively. As an explicit application of the theorem, we presented the field equations of the Horndeski's vector-tensor theory in the KSK spacetimes.


\section*{Acknowledgements}

This work is partially supported by the Scientific and Technological Research Council of Turkey (TUBITAK).


\begin{thebibliography}{EMG}

\bibitem{ggst} \.{I}. G{\" u}ll{\" u}, M. G{\" u}rses, T. \c{C}. \c{S}i\c{s}man, and B. Tekin, \textit{AdS waves as exact solutions to quadratic gravity,} Phys. Rev. D {\bf 83}, 084015 (2011).
\bibitem{gst1} M. G{\" u}rses, T. \c{C}. \c{S}i\c{s}man, and B. Tekin, \textit{New exact solutions of quadratic curvature gravity}, Phys. Rev. D {\bf 86}, 024009 (2012).
\bibitem{ghst} M. G{\" u}rses, S. Hervik, T. \c{C}. \c{S}i\c{s}man, and B. Tekin, \textit{Anti-de Sitter-wave solutions of higher derivative theories}, Phys. Rev. Lett. {\bf 111}, 101101 (2013).
\bibitem{gst2} M. G{\" u}rses, T. \c{C}. \c{S}i\c{s}man, and B. Tekin, \textit{AdS-plane wave and $pp$-wave solutions of generic gravity theories},  Phys. Rev. D {\bf 90}, 124005 (2014).
\bibitem{gst3} M. G{\" u}rses, T. \c{C}. \c{S}i\c{s}man, and B. Tekin, \textit{Gravity waves in three dimensions}, Phys. Rev. D {\bf 92}, 084016 (2015).
\bibitem{gst4} M. G{\" u}rses, T. \c{C}. \c{S}i\c{s}man, and B. Tekin, \textit{Kerr-Schild-Kundt metrics are universal}, Class. Quantum Grav. {\bf 34} 075003 (2017).

\bibitem{hervik1} A. A. Coley, G. W. Gibbons, S. Hervik, and C. N. Pope, Class. Quantum Grav. {\bf 25}, 145017 (2008).
\bibitem{hervik2} S. Hervik, V. Pravda, and A. Pravdova, \textit{Type III and N universal spacetimes}, Class. Quantum Grav. {\bf 31}, 215005 (2014).

\bibitem{pravda1} M. Ortaggio and V. Pravda, \textit{Electromagnetic fields with vanishing scalar invariants}, Classical Quantum Gravity {\bf 33},
115010 (2016).
\bibitem{pravda2} M. Ortaggio and V. Pravda, \textit{Electromagnetic fields with vanishing quantum corrections}, Phys. Lett. B {\bf 779}, 393 (2018).
\bibitem{pravda3} M. Kuchynka and M. Ortaggio, \textit{Einstein-Maxwell fields with vanishing higher-order corrections}, Phys. Rev. D {\bf 99}, 044048 (2019).
\bibitem{Kuchn} M. Kuchynka, T. Malek, V. Pravda and A. Pravdova, \textit{Almost universal spacetimes in higher-order gravity theories}, Phys. Rev. D {\bf 99}, 024043 (2019).

\bibitem{ks} R. P. Kerr and A. Schild, \textit{Some algebraically degenerate solutions of Einstein's gravitational field equations}, Proc. Symp. Appl. Math. {\bf 17}, 199 (1965); G. C. Debney, R. P. Kerr, and A. Schild, \textit{Solutions of the Einstein and Einstein-Maxwell Equations}, J. Math. Phys. {\bf 10}, 1842 (1969).
\bibitem{gg} M. G{\" u}rses and F. G{\" u}rsey, \textit{Lorentz covariant treatment of the Kerr-Schild geometry}, J. Math. Phys. {\bf 16}, 2385 (1975).

\bibitem{horn} G. W. Horndeski, \textit{Conservation of charge and the Einstein-Maxwell field equations}, J. Math. Phys. {\bf 17}, 1980 (1976).


\bibitem{gh} M. G{\" u}rses and M. Halilsoy, \textit{PP-waves in the generalized Einstein theories}, Phys. Lett. A {\bf 68}, 182 (1978).

\bibitem{ghs} M. G{\" u}rses, Y. Heydarzade, and \c{C}. \c{S}ent\"{u}rk, \textit{Kerr-Schild-Kundt metrics in generic gravity theories with modified Horndeski couplings}, Eur. Phys. J. C {\bf 81}, 1147 (2021).




\end{thebibliography}
\end{document}